\documentclass[runningheads]{llncs}
\usepackage{graphicx}
\begin{document}
\title{Influence of centrality definition and detector efficiency on the net-proton kurtosis}
%
%
\author{Sukanya Sombun\inst{1} \and
Jan  Steinheimer\inst{2} \and
Christoph Herold\inst{1} \and Ayut Limphirat\inst{1} \and Yupeng Yan\inst{1} \and Marcus Bleicher\inst{2,3,4,5} }
\authorrunning{S. Sombun et al.}
%
\institute{School of Physics and Center of Excellence in High Energy Physics $\&$ Astrophysics, Suranaree University of Technology, Nakhon Ratchasima 30000, Thailand \and
Frankfurt Institute for Advanced Studies, Ruth-Moufang-Str. 1, 60438 Frankfurt am Main, Germany\\
 \and
Institut f\"ur Theoretische Physik, Goethe Universit\"at Frankfurt, Max-von-Laue-Strasse 1, D-60438 Frankfurt am Main, Germany \and
GSI Helmholtzzentrum f\"ur Schwerionenforschung GmbH, Planckstr. 1, 64291 Darmstadt , Germany \and
John von Neumann-Institut f\"ur Computing, Forschungzentrum J\"ulich,52425 Jülich, Germany\\
\email{sukanya$\_$joy33@hotmail.com}}

\maketitle              
\begin{abstract} 
 We study the influence of the centrality definition and detector efficiency on the net-proton kurtosis for minimum bias Au+Au collisions at a beam energy of $\sqrt{s_{\mathrm{NN}}}= 7.7$ GeV by using the UrQMD model. We find that different ways of defining the centrality lead to different cumulant ratios. Moreover, we demonstrate that the kurtosis is suppressed for central collisions when a wider transverse momentum acceptance is used. Finally, the influence of a detector efficiency on the measured cumulant ratios is estimated.
 

\end{abstract}
\section{Introduction}
One aim of heavy ion collisions is to study the phase structure of the strong interaction or quantum chromodynamics (QCD) which has been widely investigated both in experiment and theory. QCD based models predict a first-order phase transition at large $\mu_{B}$ which ends in a QCD critical point (CP) \cite{stephanov1998signatures}. The QCD phase structure can be disclosed from the study of event-by-event fluctuations of conserved quantities which can be defined in the form of cumulants. Especially cumulants of sthe net-proton number are predicted to be sensitive to the presence of a QCD phase transition \cite{randrup2004spinodal,sasaki2007density} and CP \cite{stephanov2009non,herold2018broadening,nahrgang2013influence,szymanski2020net}. Higher order cumulants are particularly sensitive to a divergent correlation length close to the CP. The higher order susceptibilities of baryon number, electric charge and strangeness have been calculated theoretically in \cite{cheng2009baryon,fu2010fluctuations,chen2011statistical,karsch2011has,schaefer2012qcd,wang2012energy,zhou2012higher,rau2014conserved,fan2019probing}. Experimentally, event-by-event fluctuations of higher order cumulants of net-proton, net-pion and net-kaon number were measured by RHIC \cite{aggarwal2010higher,luo2011probing,adamczyk2014energy,adamczyk2014beam,adare2016measurement,xu2016energy} and LHC \cite{abelev2013net,rustamov2017net,ohlson2019measurements}. Hereby, a non-monotonic behaviour of higher order cumulant ratios as a function of beam energy might disclose critical behaviour. Although critical fluctuations have been widely studied, there are still difficulties in understanding the interplay of different effects and their impact on the measured observables. Some uncertainties that we are going to address in this paper using the UrQMD model are the centrality determination, the importance of volume fluctuations, the transverse momentum $(p_{T})$ acceptance, and efficiency corrections. 

\subsection{The UrQMD model}
For the present study, we use the Ultrarelativistic Quantum Molecular Dynamics (UrQMD) model to study cumulant ratios of the net-proton number distribution. The UrQMD model is a microscopic transport model which is able to simulate p+p, p+A and A+A collisions. It is based on the covariant propagation of constituent quarks and anti-quarks accompanied by mesonic and baryonic degrees of freedom, binary elastic and inelastic scattering of hadrons, resonance excitations as well as string dynamics and strangeness exchange reactions \cite{bass1998microscopic,bleicher1999relativistic,graef2014deep}. The elementary cross-sections are interpreted geometrically and are taken from experimental data \cite{patrignani2016review}.

\section{Method}\label{method}
\subsection{Calculation of cumulants}
The fluctuations of the net-proton number distribution can be characterized by the corresponding cumulants. These are calculated as
\begin{eqnarray}
C_1 &=& M  = \left\langle \mathrm{N} \right\rangle \\
C_2 &=&\sigma^2 = \left\langle (\delta \mathrm{N})^2 \right\rangle \\
C_3 &=& S \sigma^{3} =\left\langle (\delta \mathrm{N})^3 \right\rangle \\
C_4 &=& \kappa \sigma^{4} =\left\langle (\delta \mathrm{N})^4 \right\rangle - 3 \left\langle (\delta \mathrm{N})^2 \right\rangle^2 
\end{eqnarray}
 where  $(\delta \mathrm{N}) = \mathrm{N} - \left\langle \mathrm{N} \right\rangle$, and $\left\langle \mathrm{N} \right\rangle$ is the event-averaged value of the net-proton number $\mathrm{N}$.
We consider the following ratios of these cumulants, 
\begin{eqnarray} 
\frac{{{C _2}}}{{{C _1}}} &=& \frac{{\left\langle {\delta {N^2}} \right\rangle }}{{\left\langle N \right\rangle }} = \frac{{{\sigma ^2}}}{M}\\
\frac{{{C _3}}}{{{C _2}}} &=& \frac{{\left\langle {\delta {N^3}} \right\rangle }}{{\left\langle {\delta {N^2}} \right\rangle }} = S\sigma\\
\frac{{{C _4}}}{{{C _2}}} &=& \frac{{\left\langle {\delta {N^4}} \right\rangle }}{{\left\langle {\delta {N^2}} \right\rangle }} - 3\left\langle {\delta {N^2}} \right\rangle  = \kappa {\sigma ^2}
\end{eqnarray}
with combinations of mean (M), variance ($\sigma^2$), skewness ($S$) and kurtosis ($\kappa$) of measured event-by-event fluctuations. For the statistical error calculations, we use the the delta theorem \cite{luo2012error} which states that the error of cumulant ratios is proportional to a certain power of the standard deviation as: 
\begin{equation}
	error(\frac{C_{r}}{C_{2}})\propto\frac{\sigma^{r-2}}{\sqrt{n}}~.
\end{equation}
	Here,  $r$  denotes the order of the cumulant, and $n$ the number of events.
\subsection{Centrality definition}
\begin{figure}[ht!]	
\includegraphics[width=\textwidth]{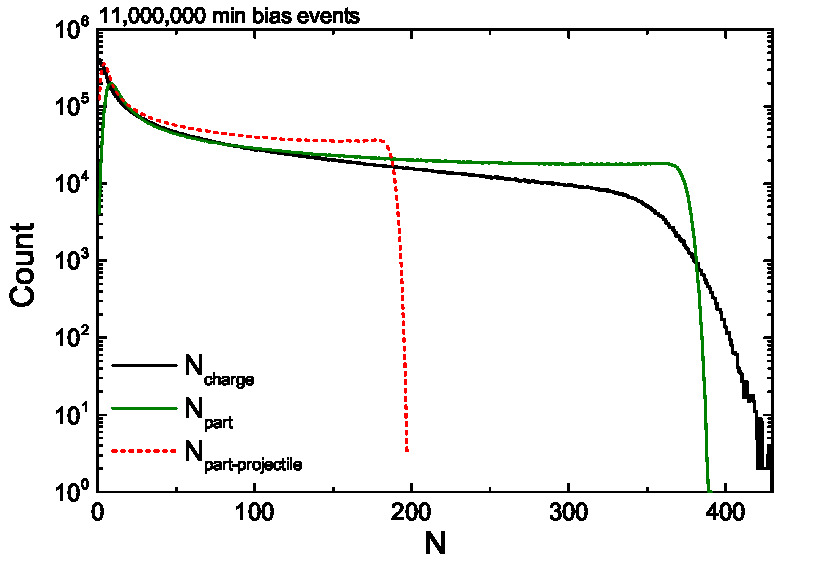}
\caption{[Color online] Distributions of $N_{charge}$, $N_{part}$ and $N_{\rm part-projectile}$ in Au+Au collision at a beam energy of $\sqrt{s_{\mathrm{NN}}}= 7.7$ GeV with impact parameter $0 \le b \le 15$~fm. Figure from \cite{sombun2017higher}.}\label{f1}
\end{figure}

In heavy-ion experiments, there is no unique definition of centrality. As the impact parameter ($b$) is experimentally not accessible, observables like the number of participants ($N_{\mathrm{part}}$) or the number of charged particles ($N_{\mathrm{charge}}$) are used. These can be related to the impact parameter e.g. by a Glauber model. Therefore, the centrality is practically determined by particle multiplicities. In our work, the centrality will be defined by $N_{\mathrm{part}}$, $N_{\mathrm{charge}}$ or the number of participants in the projectile, $N_{\mathrm{part-projectile}}$. Calculation of cumulants as a function of  $N_{\mathrm{part}}$ and $N_{\mathrm{charge}}$ is believed to minimize volume fluctuations which can have an influence on the extracted ratio of cumulants. We study the dependence of the net-proton number  on the different centrality definitions. For this, we simulate Au+Au collision at a beam energy of $\sqrt{s_{\mathrm{NN}}}=7.7$ GeV
with impact parameters in the range $0 \le b \le 15$~fm. We define the following quantities,
\begin{itemize}
\item $N_{\rm charge}$: The number of all charged particles with $|\eta|\le 1$ and $0.15<p_T<2.0$~GeV minus the number of protons and anti-protons in this specific acceptance range.
\item $N_{\rm part}$: 394 minus all spectator protons and neutrons defined by $|y|>1.5$ and $p_T<0.3$~GeV.
\item $N_{\rm part-projectile}$: 197 minus all projectile spectator protons and neutrons defined by $y>1.5$ and $p_T<0.3$~GeV.
\end{itemize}
We determine the distribution of these three different quantities as shown in figure \ref{f1}. The three different methods give different distributions. We find that the participant distribution of $N_{\rm part}$ shows a sharp cutoff at the maximum number of participants, whereas the distribution of $N_{\rm charge}$ shows a much smoother drop. The centrality is classified into 10 centrality bins of 0-10$\%$, 10-20$\%$, 20-30$\%$, 30-40$\%$, 40-50$\%$, 50-60$\%$, 60-70$\%$, 70-80$\%$, 80-90$\%$ and 90-100$\%$. We calculate the cumulants $C_n$ for a fixed $N_{\mathrm{part}}$ ($N_{\mathrm{charge}}$, $N_{\mathrm{part-projectile}}$) and then average those cumulants over all $N_{\mathrm{part}}$ in a given centrality bin. The ratios of cumulants are then determined as ratios of averages.

\section{Results}
\subsection{Dependence on centrality definition}
\begin{figure}[h]	
\includegraphics[width=\textwidth]{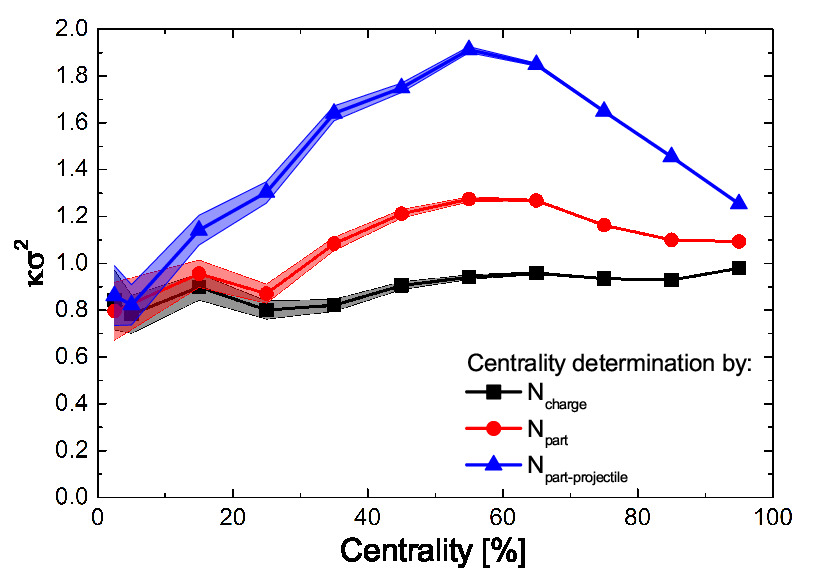}
\caption{[Color online] The kurtosis of the net-proton number as a function of centrality which is defined by three different quantities, the number of charged particles, the number of participants and the number of participants in the projectile. Figure from \cite{sombun2017higher}.}\label{f2}
\end{figure}

We first assume that the efficiency of the detector and its acceptance are perfect (100$\%$) for all particles. Figure \ref{f2} shows the result of the net-proton number kurtosis as a function of centrality for the three centrality definitions, at mid-rapidity $\left| y \right| < 0.5$ and within transverse momentum $0.4 < {p_T} < 0.8$~GeV. For the most central collisions, we find that the value of the kurtosis does not depend on the centrality definition. On the other hand, for mid-central collision, the difference between the values of $\kappa\sigma^2$ becomes larger. Moreover, the centrality defined by $N_{\mathrm{charge}}$ yields only a mild dependence of the kurtosis on the centrality because it not subject to large volume fluctuations. Thus, we use $N_{\mathrm{charge}}$ in what follows to define the centrality to investigate other effects.

\subsection{Effect of transverse momentum range}
\begin{figure}[h]	
\includegraphics[width=\textwidth]{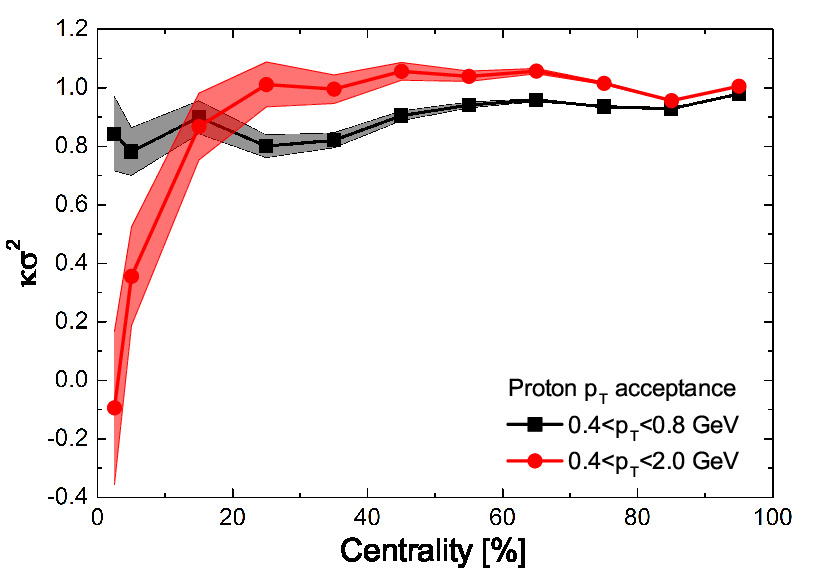}
\caption{[Color online] The kurtosis of the net-proton number as a function of centrality (defined by $N_{\rm charge}$) for two different transverse momentum $p_{T}$ ranges. Figure from \cite{sombun2017higher}.}\label{f3}
\end{figure}

We now study the kurtosis of the net-proton number as a function of centrality which is defined by $N_{\mathrm{charge}}$ for two different transverse momentum ranges, namely ($0.4 < {p_T} < 0.8$ GeV) and  ($0.4 < {p_T} < 2.0$ GeV). The result is shown in figure \ref{f3}. It can be seen that for the most central collisions, the value of the kurtosis is strongly suppressed for the larger acceptance range due to baryon number conservation. At mid-central collisions, the value of the kurtosis is larger for the wider transverse momentum range which indicates that volume fluctuations affect the extracted kurtosis.

\subsection{Effect of efficiency}
\begin{figure}[ht!]	
\includegraphics[width=\textwidth]{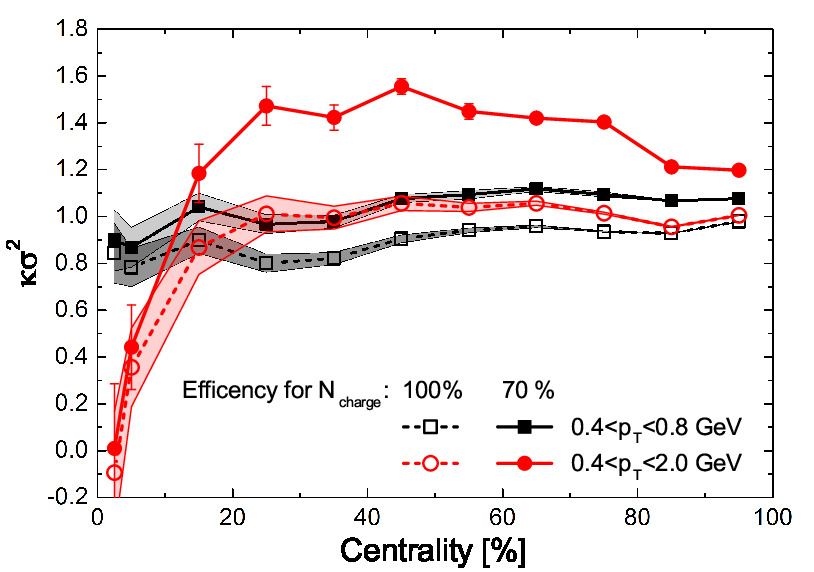}
\caption{[Color online] The kurtosis of the net-proton number in two different $p_{T}$ acceptance bins (black squares and red circles) as a function of centrality with $100\%$ (open symbols) and $70\%$ (full symbols) $N_{\rm charge}$ efficiency. Figure from \cite{sombun2017higher}.}\label{f4}
\end{figure}

In practice, particle detectors are not ideal systems but suffer from a finite particle detection efficiency. The efficiency is the number of produced particles that are recorded in the detector divided by the overall yield, see \cite{abelev2009systematic}. Therefore, detectors measuring particle number fluctuations will never perform perfectly. We now show how an efficiency below $100\%$ influences the measurement of cumulants. We begin by studying the effect of efficiency on the fluctuations. In figure \ref{f4}, we show the result of the net-proton number kurtosis in two different $p_{T}$ ranges as a function of centrality at $100\%$ and $70\%$ $N_{\rm charge}$ efficiency. It is found that the reduced (realistic) efficiency leads to an overall increase of the kurtosis.

\begin{figure}[ht!]	
\includegraphics[width=\textwidth]{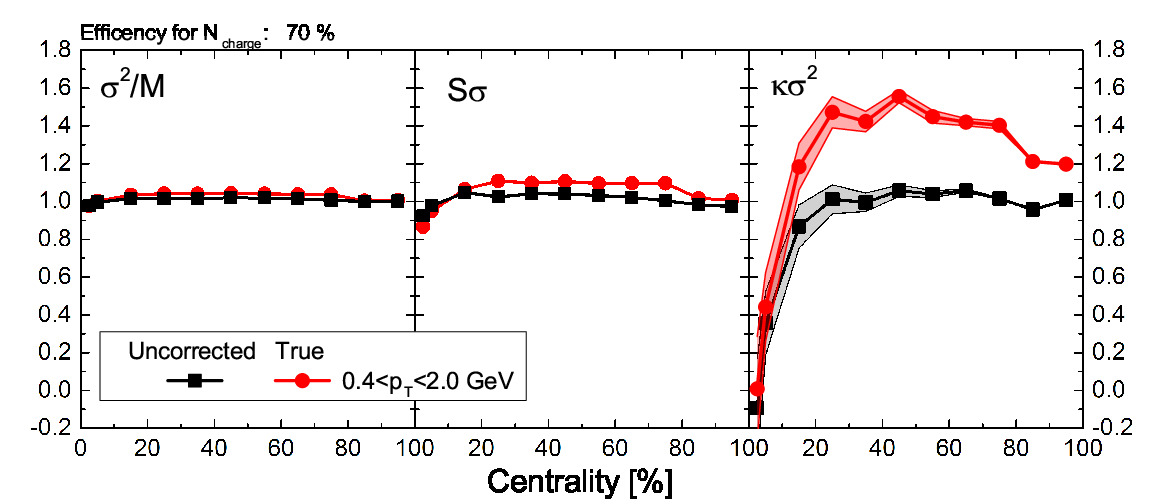}
\caption{[Color online] The result of variance, skewness and kurtosis of the net-proton number in two different $p_{T}$ acceptance bins for a 70$\%$ $N_{\rm charge}$ efficiency. We compare the result between 100$\%$ (red circle symbol) and 75$\%$ (black squares symbol) proton efficiency. Figure from \cite{sombun2017higher}.}\label{f5} 
\end{figure}

Second, we study the effect of a proton efficiency. The result of the variance, skewness and kurtosis of the net-proton number as a function of centrality defined by $70\%$ $N_{\rm charge}$ efficiency is shown in figure \ref{f5}.  We compare the results for $100\%$ and $75\%$ proton efficiency. The circles represent calculations with $100\%$ proton efficiency, the squares are $75\%$ proton efficiency. We find that the cumulant ratios of the net-proton number for the $75\%$ proton efficiency is smaller than for the 100$\%$ proton efficiency.

\section{Conclusions}
We have studied the effect of different methods for centrality determination on the measured net-proton kurtosis. We have found that different centrality definitions give different results for the kurtosis.  Using a centrality defined by $N_{\rm charge}$ reduces the effect of volume fluctuations. Moreover, we have studied the effect of two different transverse momentum ranges accepting net-protons in the measurement. We have seen that the wider transverse momentum range leads to a strongly suppressed kurtosis at central collisions. Finally, we have observed the effect of centrality determination, finding a clear impact on the value of the kurtosis.

\section{Acknowledgments}
The computational resources have been provided by the LOEWE Frankfurt Center for Scientific Computing (LOEWE-CSC) and the Center for Computer Services at SUT. This work is supported by the German Academic Exchange Service (DAAD), HIC for FAIR and the Thailand Research Fund (TRF). SS and AL acknowledge support from TRF-RGJ (PHD/0185/2558). CH, AL and YY acknowledge support from Suranaree University of Technology and the Office of the Higher Education Commission under NRU project of Thailand.

\bibliographystyle{ieeetr}
\bibliography{bibtex}

\end{document}